\documentclass[aps,pra,twocolumn,superscriptaddress,showpacs]{revtex4-1}
\usepackage{amsmath,amsfonts,amsbsy,amssymb,amsthm}
\usepackage{bbold,bbm,bm}
\usepackage{graphicx}
\usepackage{hyperref}
\usepackage{qcircuit}
\hypersetup{
  colorlinks=true,
   linkcolor=blue,
   filecolor=magenta,      
    urlcolor=cyan,
}
\usepackage{multirow}
 
\theoremstyle{plain}

\newcommand{\comment}[1]{}
\newcommand{\tr}{\text{Tr}}

\newcommand{\nn}{\nonumber\\}
\newcommand{\ket}[1]{| #1 \rangle}
\newcommand{\bra}[1]{\langle #1 |}

\newcommand{\eye}{\mathbb{1}}
\newcommand{\mS}{\mathcal{S}}
\newcommand{\mG}{\mathcal{G}}
\renewcommand{\qedsymbol}{$\blacksquare$}

\begin{document}

\title{Validating and Certifying Stabilizer States}
\author{Amir Kalev}
\email{amirk@umd.edu}
\affiliation{Joint Center for Quantum Information and Computer Science, University of Maryland, College Park, MD 20742-2420, USA}
\author{Anastasios Kyrillidis}
\email{anastasios@rice.edu}
\affiliation{Department of Computer Science, Rice University, Houston TX 77005, USA}
\author{Norbert M. Linke}
\email{linke@umd.edu}
\affiliation{Joint Quantum Institute,
University of Maryland, College Park, MD 20742-2420, USA}

\date{\today}

\begin{abstract}
We propose a measurement scheme that validates the preparation of an $n$-qubit stabilizer state. 
The scheme involves a measurement of $n$ Pauli observables, a priori  determined from the stabilizer state and which can be realized using single-qubit gates.  Based on the proposed validation scheme, we derive an explicit expression for the worst-case fidelity, i.e., the minimum fidelity between the stabilizer state and any other state consistent with the measured data. We also show that the worst-case fidelity can be certified, with high probability, using ${\cal O}(n^2)$ copies of the state. 
\end{abstract}

%\pacs{}
%\keywords{} 
\maketitle

\section{Introduction}\label{sec:intro}
Quantum tomography  is the canonical procedure for diagnosing and characterizing quantum states and operations. 
The main goal of quantum tomography is to obtain detailed information that would allow us to improve the performance of a quantum device. 
However, extracting this information from experimental data is generally a hard task, even for small-sized systems. 
In addition, as the resources for quantum tomography scale exponentially with the number of subsystems, fully characterizing the system is experimentally impractical, even for systems
with a moderate number of qubits, and under strong assumptions such as purity and unitary dynamics~\cite{gross2010quantum,kalev2015quantum,ODonnell2016,Haah2017Sample}.

Therefore, rather than focusing on error diagnosis of quantum processors, we are often concerned with the simpler question of validation: e.g., checking how close is the state of the system to a {\it target} state, where usually closeness is quantified by the fidelity figure of merit. 
Along this thread, Flammia and Liu~\cite{Flammia2011Direct}, and da Silva~{\it et al.}~\cite{daSilva2011Practical} proposed a measurement scheme, known as \emph{direct fidelity estimation}, tailored to estimate the fidelity between the (unknown) state of the system and a target state, without the need to perform full quantum tomography.

In this work, we are interested in cases where the target state is an $n$-qubit stabilizer state. 
Stabilizer states constitute an important class of states, used for teleportation-based~\cite{Gottesman1999Demonstrating} and measurement-based \cite{Raussendorf2001One} quantum computation, quantum error-correction codes~\cite{Gottesman1997Stabilizer}, and quantum self-testing~\cite{Breiner2018Parallel}. 
For such states, the result in~\cite{Flammia2011Direct} and~\cite{daSilva2011Practical} translates into a measurement of $\mathcal{O}(\epsilon^{-2}\ln(1/\delta))$ Pauli observables (picked at random from a distribution which depends on the target state), where $\epsilon$ and $\delta$ are small user-defined quantities related to the estimation error and the failure probability of the direct fidelity estimation scheme, respectively. 
Importantly, the number of observables is independent of $n$. 
In practice though, to obtain a modest, say $1\%$, estimation error with $2/3$ success probability, we would require a measurement of roughly $10^4$ Pauli observables. 

Here, we propose a practical scheme for validating an $n$-qubit stabilizer state by measuring exactly $n$ Pauli observables, which makes it relevant for small- to moderate-sized systems. 
The measurements can be realized by single-qubit gates, and the observables can be a priori chosen based on the target state. 
%The scaling with $n$ makes our protocol suitable for moderate-size systems. 
We also give an explicit, straightforward, formula for the worst-case fidelity, i.e., the minimum fidelity between the target stabilizer state and any state consistent with the measured data. 
Moreover, we show that the worst-case fidelity can be certified, with high probability, using ${\cal O}\left(n  \epsilon^{-2} \ln(1/\delta)\right)$ copies of the state of the system for each measured observable. 
When the worst-case fidelity is close to one, we prove that one can use our scheme to obtain a high-fidelity estimate of the state of the system. 

The paper is organized as follows. In the next section we set up the notation and briefly review the notions of quantum state validation and stabilizer states. In Sec.~\ref{sec:valid}, we describe a validation scheme for stabilizer states and prove its certification guarantees. Then in Sec.~\ref{sec:exp}, we apply our scheme to validate the preparation of a three-qubit GHZ state on a trapped-ion quantum computing platform~\cite{Debnath16} and superconducting quantum computing platform~\cite{IBMQ}. Finally we close with a discussion on Sec.~\ref{sec:disc}.

\section{Background and setup}\label{sec:background}
To set up the notions and notation used in this work, we first briefly review the basic idea of state validation and the theory of stabilizer states. In what follows, we refer to a quantum state \emph{validation protocol} as a measurement scheme that, in the noiseless case,  certifies with probability one that the state of the system is the target state, {\it if and only if} this is the case. 
In the presence of experimental noise, a validation scheme should provide a certification for the worst-case fidelity. 

Unlike quantum tomography, in a validation protocol we may choose the measurement scheme to depend on the target state. 
For example, given a target state $\ket{\Psi_0}$ we may consider the two-outcome positive-operator valued measure (POVM) $\left\{E_0=\ket{\Psi_0}\bra{\Psi_0},E_1=\eye-\ket{\Psi_0}\bra{\Psi_0} \right\}$ as a validation scheme. 
In this case, the worst-case fidelity is the frequency of occurrence of the outcome $E_0$. If $\ket{\Psi_0}$ is an entangled state, implementing such a POVM would involve  applying a non-local unitary to the state of the system.  In contrast, in this work we propose a validation scheme for a target $n$-qubit stabilizer state that consists of experimentally accessible measurements of Pauli observables that can be realized using single-qubit quantum gates.

In a nutshell, an $n$-qubit stabilizer state, $\ket{\Psi_0}$, is  the unique eigenstate, with eigenvalue $1$,  of $2^n$ commuting $n$-qubit Pauli operators $\{P_l\}_{l=1}^{2^n}$, where $P_l\in\{\pm 1,\pm i\}\cdot\{I,X,Y,Z\}^{\otimes n}$, and $I,X,Y,Z$ are the identity and the Pauli matrices on one qubit.  
Hereafter, we set $P_{2^n}=\eye\equiv I^{\otimes n}$. 
The set $\{P_l\}_{l=1}^{2^n}$ forms the \emph{stabilizer group} $\mS$. 
This group is generated by a set of $n$ Pauli operators (their choice is not unique) denoted here by $\mG\subset\mS$.
The stabilizer state can then be written as,
\begin{equation}\label{eq:stab}
\rho_0\equiv\ket{\Psi_0}\bra{\Psi_0}=\frac1{2^n}\prod_{P_l\in \mG}\Bigl(\eye+P_l\Bigr)=\frac1{2^n}\sum_{P_l\in \mS}P_l.
\end{equation}

\section{Validation of stabilizer states}\label{sec:valid}
As reflected by Eq.~\eqref{eq:stab},  $\ket{\Psi_0}$ is the unique eigenstate, with eigenvalue $+1$ of the $n$ generators $P_l\in \mG$. 
This, in turn, implies that the solution to the feasibility problem:
\begin{equation}
	\begin{aligned}
		& \underset{\rho \in \mathbb{C}^{2^n \times 2^n}} {\text{find}}
		& & \rho\\
		& \text{subject to}
     & & \tr(\rho P_l) =1,\;\forall P_l\in \mG,  \\
		& & & \rho \succeq 0, \; \text{Tr}(\rho) =1,
	\end{aligned} \label{prog:feas}
\end{equation}
is a singleton, $\rho_0$. 
To see that, assume that there exists a density matrix $\rho_1\neq\rho_0$ which is a solution to the feasibility problem above. 
When writing it in its eigenbasis, $\rho_1=\sum_j\lambda_j\ket{\psi_j}\bra{\psi_j}$, the feasibility conditions $\tr(\rho_1 P_l) =1$  imply that  $\sum_j\lambda_j\bra{\psi_j}P_l\ket{\psi_j}=1, $ $\forall P_l\in \mG$. 
Since the $P_l$'s have eigenvalues $\pm1$ we have $-1\leq\bra{\psi_j}P_l\ket{\psi_j}\leq1$, and due to the positivity of the $\lambda_j$'s we obtain the inequality   $-1=-\sum_j\lambda_j\leq\sum_j\lambda_j\bra{\psi_j}P_l\ket{\psi_j}\leq\sum_j\lambda_j=1$. 
The upper bound is obtained when  $\bra{\psi_j}P_l\ket{\psi_j}=1, ~\forall P_l\in\mG$ and $\forall ~\ket{\psi_j}$ in the eigenbasis of $\rho_1$. 
But since, by definition, $\ket{\Psi_0}$ is the unique pure state for which  $\bra{\Psi_0}P_l\ket{\Psi_0}=1$, $\forall P_l\in\mG$, we obtain $\rho_1=\rho_0$ in contradiction to our initial assumption.  

Note, however, that while $\rho_0$ is the unique solution to program~\eqref{prog:feas}, there are infinitely many Hermitian  matrices with negative eigenvalues (i.e., that do not satisfy the constraint $\rho \succeq 0$ above) for which $\tr(\rho P_l) =1,\;\forall P_l\in \mG$. 
All of which have the structure $\frac1{2^n}\sum_{P_l\in \mS}P_l +\sum_{P_l\notin \mS}c_l P_l$ (considering those with trace 1) for some real numbers $c_l$. 
Therefore, constraining on density matrices in~\eqref{prog:feas} is crucial to obtain a singleton solution.

Hence, in the absence of noise, the only quantum state that is consistent with the noiseless ``data'' $\tr(\rho P_l) =1,\;\forall P_l\in \mG$, is the stabilizer state $\ket{\Psi_0}$.  
Therefore, given a target $n$-qubit stabilizer state, we can consider the  measurement of the expectation values of a set of $n$ generators as a validation scheme. 
Since the generators are mutually commuting, there is, in principle, a measurement scheme to measure them simultaneously.

Consider, for example, the case where the target state is the $n$-qubit GHZ stabilizer state $\ket{\textrm{GHZ}_n}=\frac1{\sqrt 2}(\ket{0}^{\otimes n}+\ket{1}^{\otimes n})$, where $\{\ket{0},\ket{1}\}$ are the eigenbasis of the $Z$ Pauli matrix. 
To validate that this is indeed the state of the system, we have the freedom to choose a specific set of generators that can be measured.  
A convenient choice of stabilizer generators are the $n$ Pauli observables, $X^{\otimes n}$ and $Z_k \otimes Z_{k+1}$ for $k=1,\ldots,n-1$. 
We can measure the expectation values of these observables with two simple experimental setups, for any $n$. In the first setup, we measure all the qubits in the $X$ basis, while in the second setup we measure all the qubits in the computational basis, $Z$. The expectation values of the generators above can be calculated from the experimental results. Note that these measurements only involve single-qubit gates. 

We note that, since in the noiseless case the feasibility program~\eqref{prog:feas} has a unique solution, due to convexity, in the presence of small experimental noise, the argument solution to:
\begin{equation}
	\begin{aligned}
		& \underset{\rho \in \mathbb{C}^{2^n \times 2^n}} {\text{minimize}}
		& & \mathcal{C}(\rho)\\
		& \text{subject to}
     & & \tr(\rho P_l) =1-\varepsilon_l,\;\forall P_l\in \mG,  \\
		& & & \rho \succeq 0, \; \text{Tr}(\rho) =1,
	\end{aligned} \label{prog:conv}
\end{equation}
where $\mathcal{C}(\rho)$ is a convex function of $\rho$, and $\varepsilon_l\geq0$ captures experimental errors, is guaranteed to be close,  in fidelity, to $\rho_0$ (which is assumed to be close to the state of the system).  This follows directly from our result,  Proposition~1 below, which implies that,  in the case of small $\varepsilon_l$'s, the argument solution to program~\eqref{prog:conv} have fidelity at least  $1-\frac1{2}\sum_{l:P_l\in \mG} \varepsilon_l$ with the target state of the system.   

Next, we show that the proposed validation scheme leads to an experimentally-useful lower bound on the worst-case fidelity to the target stabilizer state. 
Since $\rho_0$ is a pure state, the fidelity between $\rho_0$ and any other state $\rho$ is $F=\tr(\rho_0\rho)$. 
Thus, given the experimental data for the expectation values of the generators, $\tilde{\mu}_1,\ldots,\tilde{\mu}_n$,  to find a lower bound on the worst-case fidelity, we can solve the convex program:
\begin{equation}
	\begin{aligned}
		& \underset{\rho \in \mathbb{C}^{2^n \times 2^n}} {\text{minimize}}
		& & \tr(\rho_0\rho)\\
		& \text{subject to}
     & & \tr(\rho P_l) =\tilde{\mu}_l,\;\forall P_l\in \mG,  \\
		& & & \rho \succeq 0, \;  \text{Tr}(\rho) =1.
	\end{aligned} \label{prog:fid}
\end{equation}

\medskip
{\it Proposition 1:} If $\sum_{l:P_l\in \mG}\frac{1-\tilde{\mu}_l}{2}\leq 1$, the solution to program~\eqref{prog:fid} is 
\begin{equation}\label{eq:fid_min}
F_\text{min}=1-\sum_{l:P_l\in \mG}\frac{1-\tilde{\mu}_l}{2}.%\frac1{2}\Bigl(\sum_{l=0}^{n-1} \tilde{\mu}_l - (n-2)\Bigr). 
\end{equation}
Otherwise, if $\sum_{l:P_l\in \mG}\frac{1-\tilde{\mu}_l}{2}> 1$, the solution to program~\eqref{prog:fid} is $F_\text{min}=0$.

\medskip
{\it Proof:} Let us order the stabilizer operators such that $P_0,\ldots,P_{n-1}$ are the measured generators. 
Then, an $n$-qubit density matrix that is consistent with the data is given by
\begin{equation}\label{eq:estimate1}
\rho=\frac1{2^n} \left(\eye+\sum_{l=0}^{n-1} \tilde{\mu}_l P_l +\sum_{l=n}^{2^n -1} x_l P_l\right).
\end{equation}
We can add more terms to $\rho$, in the subspace that lies outside the stabilizer group, and due to the orthogonality property of Pauli observables ($\tr(P_i P_j)\propto\delta_{i,j}$) the resulting state would be still consistent with the measured data. 
However, for the same reason, adding such terms will not change the fidelity with $\ket{\Psi_0}$. 
Therefore, for the purpose of the proof, without loss of generality, we can consider  the density matrix of~\eqref{eq:estimate1} as the most  general state consistent with the data.  

Since the Pauli observables in the stabilizer group are mutually commuting, $\rho_0=\ket{\Psi_0}\bra{\Psi_0}$ and $\rho$ of Eq.~\eqref{eq:estimate1} are commuting, thus can be diagonalized simultaneously. 
Therefore, it is convenient to re-write $\rho$ of Eq.~\eqref{eq:estimate1} as
\begin{equation}\label{eq:estimate2}
\rho=\lambda_0\ket{\Psi_0}\bra{\Psi_0}+\sum_{k=1}^{2^n-1}\lambda_k \ket{\Psi_k}\bra{\Psi_k}.
\end{equation}
where $\left\{\ket{\Psi_k} \right\}_{k=0}^{2^n-1}$ forms an orthonormal basis for the Hilbert space of $n$ qubits, such that for all $k$ and $P_l\notin\mS$, $\bra{\Psi_k}P_l\ket{\Psi_k}=0$. 
In the form of Eq.~\eqref{eq:estimate2}, it is clear that program~\eqref{prog:fid} minimizes the eigenvalue $\lambda_0$, or equivalently maximizes $\sum_{k=1}^{2^n-1}\lambda_k$, while keeping all the eigenvalues non-negative.

At this point, we have the freedom to choose the basis vectors $\{\ket{\Psi_k}\}_{k=1}^{2^n-1}$.  
A suitable choice is to define these vectors through the projection operators $\frac{\eye\pm P_l}{2}$ associated with the stabilizer generators $P_l$, $l=0,\ldots,n-1$, that is, 
\begin{equation}\label{eq:basis}
\ket{\Psi_k}\bra{\Psi_k} = \prod_{l=0}^{n-1}\frac1{2}\Bigl(\eye+(-1)^{b_{l}^{(k)}}P_{l}\Bigr),
\end{equation}
for $k=0,1,\ldots,2^n-1$, and  $b_{l}^{(k)}=\{0,1\}$ is the $l$-th bit in the binary representation of $k$, $k=\sum_{l=0}^{n-1} b_l^{(k)} 2^l$. Writing $\rho$ in this basis we obtain 
\begin{equation}\label{eq:estimate3}
\rho=\sum_{k=0}^{2^n-1}\lambda_k  \prod_{l=0}^{n-1}\frac1{2}\Bigl(\eye+(-1)^{b_{l}^{(k)}}P_{l}\Bigr).
\end{equation}

The condition that $\rho$ should be consistent with the data   implies that $\tr(\frac{\eye-P_l}{2}\rho)=\frac{1-\tilde{\mu}_l}{2}$ for all generators $P_l\in \mG$. 
Using the expression for $\rho$ of Eq.~\eqref{eq:estimate3} together with the relations $\frac{\eye-P_l}{2}\frac{\eye+P_l}{2}=0$, and $\bigr(\frac{\eye-P_l}{2}\bigl)^2=\frac{\eye-P_l}{2}$,  yields the set of $n$ constrains
\begin{equation}\label{eq:constraints1}
\tr\Bigl(\frac{\eye-P_{l}}{2}\rho\Bigr)=\frac{1-\tilde{\mu}_l}{2}=\sum_{k:b_l^{(k)}=1}\lambda_{k},
\end{equation}
for $l=0,1,\ldots,n-1$. Since the eigenvalues of the Pauli observables are $\pm1$, the experimental values $\tilde{\mu}_l\in[-1,1]$ and $\frac{1-\tilde{\mu}_l}{2}\geq0$. Importantly, since for $k=0$ $b_l^{(0)}=0$ for all $l$, the set of equations~\eqref{eq:constraints1} do not contain $\lambda_0$. Moreover, the right-hand-side of the set of equations~\eqref{eq:constraints1}, for $l=0,1,\ldots,n-1$,  contain all of the  eigenvalues $\lambda_k$, $k=1\ldots,2^n-1$, with various multiplicities. Therefore, summing Eq.~\eqref{eq:constraints1} over $l$, we can write  
\begin{equation}\label{eq:constraints2}
0\leq\sum_{l=0}^{n-1}\frac{1-\tilde{\mu}_l}{2}=\sum_{k=1}^{2^{n}-1}\lambda_{k}+\Lambda(\{\lambda_{k}\}),
\end{equation}
where $\Lambda(\{\lambda_{k}\})$ denotes the sum of all the terms not in $\sum_{k=1}^{2^{n}-1}\lambda_{k}$ (its structure is not important for the proof). Since  $\forall k$ $\lambda_{k}\geq0$, $\Lambda(\{\lambda_{k}\})$ must be non-negative as well. Let us assume that $\sum_{l=0}^{n-1}\frac{1-\tilde{\mu}_l}{2}\leq1$. Therefore,  according to Eq.~\eqref{eq:constraints2}, the maximal value of the sum $\sum_{k=1}^{2^{n}-1}\lambda_{k}$ is obtained for $\Lambda(\{\lambda_{k}\})=0$, that is, $\sum_{k=1}^{2^{n}-1}\lambda_{k}=\sum_{l=0}^{n-1}\frac{1-\tilde{\mu}_l}{2}$.  In this case, the minimal value of $\lambda_0$ is
 \begin{equation}\label{eq:lambda0}
\lambda_0=1-\sum_{k=1}^{2^n-1}\lambda_k=1-\sum_{l=0}^{n-1}\frac{1-\tilde{\mu}_l}{2}\geq0.
\end{equation}
If, on the other hand, $\sum_{l=0}^{n-1}\frac{1-\tilde{\mu}_l}{2}=1+\Delta$, for some $\Delta>0$, then from Eq.~\eqref{eq:constraints2}, it is clear that  the maximal value of the sum $\sum_{k=1}^{2^{n}-1}\lambda_{k}$ is obtained when $\Lambda(\{\lambda_{k}\})=\Delta$, i.e., when $\sum_{k=1}^{2^{n}-1}\lambda_{k}=1$. In this case $\lambda_0=0$ and, thus, the worst-case fidelity with the target state is zero.
\hfill \qedsymbol

\medskip
We note that the eigenvalue $\lambda_{2^l}$ appears only in the $l$-th equation of~\eqref{eq:constraints1}. Therefore, for the case where $\sum_{l=0}^{n-1}\frac{1-\tilde{\mu}_l}{2}\leq 1$ a valid solution of Eq.~\eqref{eq:constraints2} is given by $\lambda_{2^l}=\frac{1-\tilde{\mu}_l}{2}\geq0$, and $\lambda_{k}=0$ for all other values $k\neq0$.  This implies that, in this case, one density matrix that minimizes program~\eqref{prog:fid} is given by
 \begin{align}\label{eq:estimate4}
\hat\rho&=\Bigl(1-\sum_{l=0}^{n-1}\frac{1-\tilde{\mu}_l}{2}\Bigr)\ket{\Psi_0}\bra{\Psi_0}\nn&+\sum_{l=0}^{n-1}\Bigl(\frac{1-\tilde{\mu}_l}{2}\Bigr) \Bigl(\frac{\eye-P_l}{2}\Bigr)\prod_{\substack{P_j\in \mG \\ j\neq l}}\Bigl(\frac{\eye+P_j}{2}\Bigr).
\end{align}
The state $\hat\rho$ is a good estimate for the state of the system when the fidelity is very close to 1, i.e., when $\tilde{\mu}_l=1-\varepsilon_l$ for small $\varepsilon_l$.

Next, we provide a certification for the worst-case fidelity. Since this certification is of interest for large values of $F_\text{min}$, we will implicitly assume that $\sum_{l=0}^{n-1}\frac{1-\tilde{\mu}_l}{2}\leq1$.

\medskip
{\it Proposition 2:} Fix the parameters $\epsilon>0$ and $\delta>0$, and use $m_l=\lceil{\frac{n^2\ln(2/\delta)}{2\epsilon^2}}\rceil$ copies of the state of the system, $\varrho$, to measure the Pauli generator $P_l$,  for  $l=0,1,\ldots,n-1$. Then, with probability at least $1-\delta$, the fidelity $F_\text{min}$ lies in the range $[\bar{F}_\text{min}-\frac{\epsilon}{2},\bar{F}_\text{min}+\frac{\epsilon}{2}]$, where $\bar{F}_\text{min}$ is obtained by replacing $\tilde{\mu}_l$ with $\mu_l=\tr(\varrho  P_l)$ in Eq.~\eqref{eq:fid_min}.

\medskip 
{\it Proof:}  Let  $y_{j}^{(l)}\in\{-1,+1\}$ be the outcome of measuring the generator $P_l$ on $j$-th copy of $\varrho$. The empirical expectation value of $P_l$ is given by  $\tilde{\mu}_l=\frac1{m_l}\sum_{j=1}^{m_l} y_{j}^{(l)}$.  By Hoeffding's inequality applied to Bernoulli experiment, the probability that $\tilde{\mu}_l$ is $\epsilon$-close to its mean $\mu_l=\tr(\varrho P_l)$ is~\cite{shalev2014}
\begin{equation}\label{eq:concen}
P\left(\left\vert \tilde{\mu}_l-\mu_l\right\vert\leq\epsilon\right)\geq1-2e^{-2m_l\epsilon^2}.
\end{equation}
Therefore, taking $m_l=\lceil{\frac{\ln(2/\delta)}{2\epsilon^2}}\rceil$ $\forall l$,  with probability at least $1-\delta$, the data $\tilde{\mu}_l$  lies in the range $[\mu_l-\epsilon,
\mu_l+\epsilon]$. Thus, with probability at least $1-\delta$, we have
\begin{align}\label{eq:fid_min2}
\bar{F}_\text{min}-\tfrac{n\epsilon}{2} &= 1-\sum_{l=0}^{n-1}\tfrac{1-\mu_l+\epsilon}{2}\leq F_\text{min} \nonumber \\ &\leq 1-\sum_{l=0}^{n-1}\tfrac{1-\mu_l-\epsilon}{2} =\bar{F}_\text{min} + \tfrac{n\epsilon}{2}.
\end{align}
Taking $\epsilon\rightarrow \epsilon/n$ completes the proof.
\hfill \qedsymbol

\medskip 
Following Proposition~1 and Proposition~2 we obtain the following corollary: 

\medskip 
{\it Corollary 3:} Let $\varrho$ be the state of the system, and let $m_l=\lceil{\frac{n^2\ln(2/\delta)}{2\epsilon^2}}\rceil$ $\forall l$. Then, with probability at least $1-\delta$,   the fidelity between $\varrho$ and the target stabilizer state $\ket{\Psi_0}$ is larger than $F_\text{min}-\frac{\epsilon}{2}$.

\medskip 
{\it Proof:}  $\bar{F}_\text{min}$ is the solution for program~\eqref{prog:fid}, i.e., the minimum fidelity with $\ket{\Psi_0}$, when we have access to the noiseless data $\mu_l=\tr(\varrho P_l)$.  
Therefore, the fidelity of the state of the system $\varrho$ and $\ket{\Psi_0}$ is necessarily larger than (or equals to)  $\bar{F}_\text{min}$. 
On the other hand, from the right-hand-side of inequality~\eqref{eq:fid_min2}, with probability $1-\delta$ we can bound  $\bar{F}_\text{min}$  from below by $F_\text{min}-n\frac{\epsilon}{2}$. 
Taking $\epsilon\rightarrow \epsilon/n$ completes the proof.
\hfill \qedsymbol

\medskip
We note that since in the proposed procedure we should measure $n$ stabilizer generators, the total sample complexity of our certification protocol is $\lceil{\frac{n^3\ln(2/\delta)}{2\epsilon^2}}\rceil$. 
For comparison, Gottesman~\cite{Gottesman2008Identifying} and Montanaro~\cite{Montanaro2017Learning}  have showed that stabilizer states  can be identified using only ${\cal O}(n)$ copies of the state. However, these methods require entangled measurements, while the method proposed in this work uses single-qubit gates. Moreover, the direct fidelity estimation scheme of~\cite{Flammia2011Direct} and~\cite{daSilva2011Practical} requires ${\cal O}\left(\text{poly}\left( \epsilon^{-1}, \ln(1/\delta)\right)\right)$ copies of the states in total for certification, independent of $n$. But, as was discussed above, their scheme is favorable for very large systems, while our scheme is applicable for moderate-size systems.

Hoeffding's inequality, used above, does not take into account the information about the variance of the outcome's distribution. 
Including this information, e.g., by using Bernstein's inequality, can further improve the sample complexity of our scheme. 
Bernstein's inequality states that given independent random variable $a_1,a_2,\ldots,a_m$ with mean $\mu$ and variance $\sigma^2$,  the probability that $\tilde\mu=\frac1{m}\sum_j a_j$ is $\epsilon$-close to $\mu$ is:
\begin{equation}\label{eq:concen2}
P\left(\left\vert \tilde{\mu}-\mu\right\vert<\epsilon\right)>1-2e^{-\frac{m\epsilon^2}{2(\sigma^2+\epsilon/3)}}.
\end{equation}
When $\sigma^2\lesssim \epsilon$,  we obtain a tighter lower bound than of Eq.~\eqref{eq:concen} which behaves like $1-2e^{-m\epsilon}$ instead of the $1-2e^{-m\epsilon^2}$, i.e., in this case, the number of samples (per observable), $m$, scale as $\frac{n\ln(2/\delta)}{\epsilon}$ rather than $\frac{n^2\ln(2/\delta)}{\epsilon^2}$. 
This can be useful for our purpose when the state of the system is close to the target stabilizer state, since then we expect the measurement outcomes of the $n$ Pauli observables to be narrowly distributed. 
(The measurement outcomes for the target stabilizer state have zero variance.) 
However, in our case, since the actual state of the system $\varrho$ is unknown, the variance of the distribution of each measured observable $(1-\tr(\varrho P_l)^2)^2$ is unknown, and in practice, we cannot use the Bernstein's inequality as stated above. 
Therefore, instead, we will use an empirical Bernstein's bound developed by~\cite{Dagum2000Optimal,Mnih2008Empirical}, which uses a variance calculated from the data, $\tilde{\sigma}^2$, instead of $\sigma^2$. 
The main idea of the  empirical Bernstein's bound of~\cite{Dagum2000Optimal,Mnih2008Empirical} is to use an online algorithm, called \emph{EBStop}, which decides when to stop taking data, so that the $\tilde{\sigma}^2$ is, with high probability, an upper bound for $\sigma^2$. 
The algorithm should be executed, in our case, for each one of the $n$ measured generators. We refer the reader to~\cite{Mnih2008Empirical} for details about the algorithm.
It was proven~\cite{Mnih2008Empirical,Mnih2008Efficient} that it takes $m={\cal O}(\epsilon^{-1}\ln\frac{1}{\delta})$ samples for the above algorithm to stop and to assure the desire convergence of $\tilde{\mu}$, with probability at least $1-\delta$.  Therefore, by taking $\epsilon\rightarrow\epsilon/n$ we find the following result:

\medskip
{\it Corollary 4:} Let $\varrho$ be the state of the system, and let $m_l={\cal O}(\frac{n\ln(1/\delta)}{\epsilon})$ $\forall l=0,1,\ldots,n-1$. Then, with probability at least $1-\delta$,   the fidelity between $\varrho$ and the target stabilizer state $\ket{\Psi_0}$ is larger than $F_\text{min}-\frac{\epsilon}{2}$.\\

\section{Experiments}\label{sec:exp}
We apply the scheme described
above to certify the preparation of a three-qubit GHZ state, $\ket{\textrm{GHZ}_3}=\frac1{\sqrt 2}(\ket{0}^{\otimes 3}+\ket{1}^{\otimes 3})$,  on two different experimental platforms. The first platform is a 5-qubit quantum computing system based on trapped Ytterbium ions with individual laser beam addressing~\cite{Debnath16}. There, the state preparation consists of two native entangling XX-gates on qubits 1-2 and 2-3, followed by Hadamard gates on all three as shown in Fig.~\ref{fig:ioncircuit}. The XX-gate is experimentally realized using the M\o{}lmer-S\o{}rensen scheme~\cite{Molmer99, Solano99, Milburn00} combined with laser-pulse shaping to control the motional modes in a multi-ion chain~\cite{Choi14}. The gate operation is given in Eq.~\ref{eqn:XXGateChi}. The parameter $\chi$ can be varied continuously, $0<\chi \leq \frac{\pi}{4}$ and is set to $\frac{\pi}{4}$ in this case for a maximally-entangling gates. Single qubit gates, which are needed to create the Hadamard operations, are generated by driving resonant Rabi rotations of defined phase, amplitude, and duration. $Z$-rotations form an exception as they are applied as phase advances on the classical controllers.
\begin{equation} \label{eqn:XXGateChi}
XX(\chi)=\begin{pmatrix}
\cos(\chi) & 0 & 0 & \hspace{-2mm}-i \sin(\chi)\\[0.3em]
0 & \hspace{-1mm}\cos(\chi) & \hspace{-3mm}-i \sin(\chi) & 0\\[0.3em]
0 & \hspace{-3mm}-i \sin(\chi) & \hspace{-1mm}\cos(\chi) & 0\\[0.3em]
-i \sin(\chi) & 0 & 0 & \cos(\chi)
\end{pmatrix}
\end{equation}
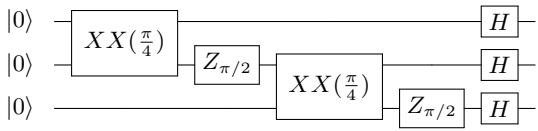
\begin{figure}[!t]
\hspace{1em}\Qcircuit @C=0.7em @R=0.2em @!R { % @C=1em @R=0.7em @!R {
%%ion 1
|0\rangle & & & \multigate{1}{XX(\frac{\pi}{4})}	& \qw & \qw	& \qw & \gate{H} & \qw \\
%% ion 2
|0\rangle & & & \ghost{XX(\frac{\pi}{4})} 			&\gate{Z_{\pi/2}} & \multigate{1}{XX(\frac{\pi}{4})} &\qw	& \gate{H} & \qw \\
%% ion 3
|0\rangle & & & \qw															&\qw						& \ghost{XX(\frac{\pi}{4})} & \gate{Z_{\pi/2}} & \gate{H} & \qw
}
\caption{Circuit used to generate the 3-qubit GHZ state in the trapped ion system. The XX-gate action is given in Eq.~\ref{eqn:XXGateChi}.}
\label{fig:ioncircuit}
\end{figure}
\begin{figure}[!t]
\hspace{1em}\Qcircuit @C=0.7em @R=0.2em @!R { % @C=1em @R=0.7em @!R {
%%ion 1
|0\rangle & & &\qw &\qw &\targ&\qw \\
%% ion 2
|0\rangle & & &\qw  & \targ &\qw &\qw \\
%% ion 3
|0\rangle & & & \gate{H}& \ctrl{-1}&\ctrl{-2}&\qw
}
\caption{Circuit used to generate the 3-qubit GHZ state on IBM's superconducting quantum computation platform.}
\label{fig:ibmcircuit}
\end{figure}
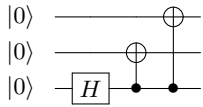
The experimental data consists of $11000$ samples from the distribution associated with measuring each qubit in the $Z$ basis (the computational basis), and $11000$ samples from the distribution associated with measuring each qubit in the $X$ basis (the Hadamard basis). The latter is achieved by applying another round of Hadamard gates before a measurement in the computational basis. The data is corrected for readout errors which are characterized independently by measuring the state-to-state transfer matrix in a series of reference experiments~\cite{Debnath16}.

For comparison, we apply the validation scheme described
above to validate a three-qubit GHZ state prepared on a IBM's superconducting quantum computation platform. We used the 5-qubit Tenerife processor~\cite{IBMQ}. The three qubits, labelled 0,1, and 2, were prepared in a  GHZ state by applying the Hadamard transformation to qubit 2, followed by CNOT operations between qubits 2 and 1 and between qubits 2 and 0, see  Fig.~\ref{fig:ibmcircuit}. Here as well, the experimental data consists of $8192$ samples from the distribution associated with measuring each qubit in the $Z$ basis, and $8192$ samples from the distribution associated with measuring each qubit in the $X$ basis. The latter is obtained by applying the Hadamard transformation to each qubit followed by a measurement in the computational basis.  
\begin{figure}[!t]
\centering
\includegraphics[width=0.9\columnwidth]{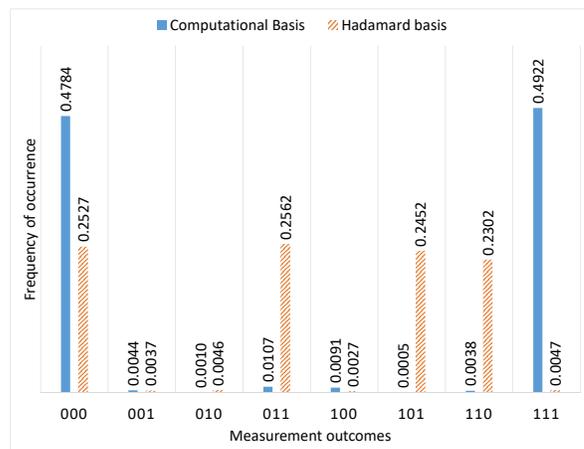}
\caption{{\bf Experimental results for trapped-ion platform.}}\label{fig:ion}
\end{figure}
\begin{figure}[!t]
\centering
\includegraphics[width=1\columnwidth]{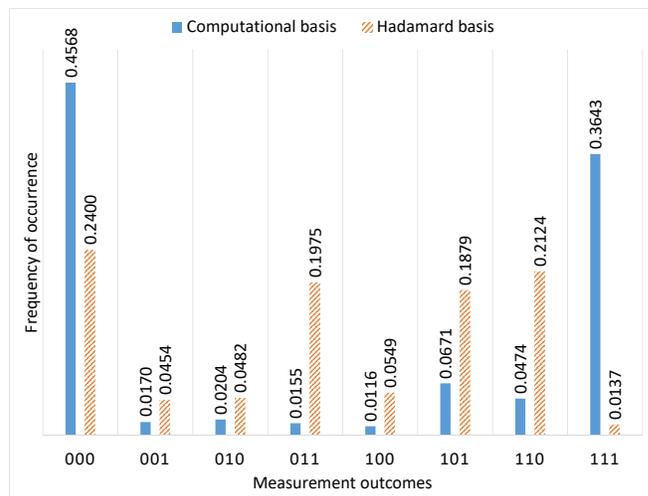}
\caption{{\bf Experimental results for IBM Q Tenerife processor.} }\label{fig:ibm}
\end{figure}

In Figs.~\ref{fig:ion} and~\ref{fig:ibm} we show the results of the two experiments in terms of the occurrence frequencies of the computational basis states.  Based on these results, using Corollary~4, we calculate for the two experiments the lower bound on fidelity between the prepared state and the three-qubit GHZ state, $\ket{\textrm{GHZ}_3}$, for various certificate values. These  are given in Table~\ref{tbl:results}. For the ion-trap (IBM) experiment we obtain these quantities by calculating the expected values of the three generators $Z_1\otimes Z_2$, $Z_2\otimes Z_3$ ($Z_1\otimes Z_3$), and $X^{\otimes3}$. For the observed data, these choice of generators maximize the lower-bound on the fidelity.
\begin{table}[h]
    \begin{tabular}{ c| c| c |}
  \cline{2-3}
& \multicolumn{2}{ c| }{Fidelity lower bound} \\ \cline{1-3}
 \multicolumn{1}{ |c| }{$p_\text{conf}$} & Ion-trap & IBM \\ \cline{1-3}
   \multicolumn{1}{ |c| }{0.999} & 0.905 & 0.575 \\\cline{1-3}
   \multicolumn{1}{ |c| }{0.99} &0.913 & 0.585 \\ \cline{1-3}
  \multicolumn{1}{ |c| }{0.9} & 0.923 & 0.596 \\  \cline{1-3}
    \end{tabular}
\caption{Based on the result shown in Figs.~\ref{fig:ion} and~\ref{fig:ibm}, we conclude that in these two experiments, with probability larger than $p_\text{conf}$, the state of the system has fidelity higher than the tabulated value with the three-qubit GHZ state.}
\label{tbl:results}
\end{table}

\section{Related work and discussion}\label{sec:disc}
The results in Table \ref{tbl:results} yield a clear difference between the lower fidelity bounds of the two quantum devices investigated. Other comparisons between these platforms have shown only small differences in performance for applications where the limited connectivity of the superconducting system did not have an impact \cite{LinkePNAS2017}. Consistent with these previous findings, the measurement outcomes on both systems represent the target distributions reasonably well, as the average probability for measuring one of the desired states is only about $15\%$ lower on the IBM machine, see Figs.~\ref{fig:ion} and~\ref{fig:ibm}. What our analysis illustrates is that the certified lower-bound fidelity metric punishes these small differences as they disproportionately increase the overlap of the measurement outcomes with other potential distributions consistent with non-GHZ states, and hence resulting in much lower fidelity bounds on the IBM results. Measuring Pauli observables outside a specific set of generators, may help to narrow down the set of non-GHZ states that are consistent with the measurement data, and can  increase the lower bound on the fidelity.  

The protocol we presented does not deal with measurement errors, and assumes that we are able to measure exactly the desired set of generators. 
In a real experimental setup, there could be a bias (either random or systematic) in the measured observables, which can decrease the lower bound on the fidelity. 
For example, consider a systematic bias in the measurement of a single qubit, where instead of measuring the projection onto  $\{\ket{0}\bra{0},\ket{1}\bra{1}\}$, we are, in practice, measuring the POVM $ \{(1-\eta) \ket{0}\bra{0}+\eta\ket{1}\bra{1}, (1-\eta) \ket{1}\bra{1}+\eta\ket{0}\bra{0}\}$ (in the $x$, $y$, and $z$ directions), for small  $\eta$. 
Then the lower bound on the fidelity decreases by $\mathcal{O}(n^2\eta)$ when the state of the system is close to the stabilizer state. 
It is, therefore, important to devise state-preparation validation schemes that are robust to measurement errors. We leave that for future work.

Finally, we wish to mention the recent work by Rocchetto~\cite{Rocchetto2017Stabiliser}, who showed that stabilizer states are efficiently probably approximately correct (PAC)-learnable. 
Based on the results of Aaronson~\cite{Aaronson2007learnability}, Rocchetto showed that stabilizer states can be learned,  in the  computational learning theory sense, with only  $\mathcal{O}(n)$ copies of the state, and proposed  a learning procedure which involves an optimization problem that can be solved on classical computer in polynomial time. 
The PAC-learnability of stabilizer (GHZ) states was also demonstrated experimentally in~\cite{Rocchetto2017Experimental}. 
The goal in the PAC-learning methodology is to provide a model for the state that produces (with high probability) good predictions of future experimental results. Nevertheless, even in the case of small experimental noise, in general, the learned state may have poor fidelity with the actual state of the system (which is assumed to be closed to the target state). This is in contrast to the validation method presented here which guarantees that the estimated state has high fidelity with the state of the system in the case of small experimental noise. In fact, in the aforementioned experiment of~\cite{Rocchetto2017Experimental}, the learned state of the system was shown to have a good fidelity with the target GHZ state. 
The high fidelity with the target state in~\cite{Rocchetto2017Experimental} can be understood from the point of view of the validation scheme presented here. Specifically, the measurement distribution in~\cite{Rocchetto2017Experimental} was chosen  to be uniform over the stabilizer group of the GHZ state, and therefore dependent on the target state. The authors reported that measuring roughly $1.2n$ stabilizer Pauli observables allowed them to obtain a high fidelity estimation of the target state by solving a convex program. Indeed, measuring slightly more than $n$ stabilizers assures that, with high probability, $n$ of them are independent, hence, form a generator group. This, as was discussed above, allows a good reconstruction of the stabilizer state.

\section*{Acknowledgements}
AK would like to thank Yi-Kai Liu for stimulating discussions and acknowledges the support from the US Department of Defense.

\bibliographystyle{apsrev}
\bibliography{biblio} 
\end{document}